\documentclass[journal,final]{IEEEtran}
\usepackage{array}
\usepackage[comma, numbers, sort&compress]{natbib}
\usepackage{xcolor}
\usepackage{url}
\usepackage{amsmath}
\usepackage{amsfonts}
\usepackage{amssymb}
\usepackage{booktabs}
\usepackage{graphicx}       
\usepackage{multirow}
\usepackage{nicefrac}       
\usepackage{microtype}      

\begin{document}
\title{Domain Fingerprints for No-reference Image Quality Assessment}

\author{Weihao~Xia, Yujiu~Yang,~\IEEEmembership{Member,~IEEE}, Jing-Hao~Xue, ~\IEEEmembership{Member,~IEEE}, Jing~Xiao,~\IEEEmembership{Member,~IEEE}

\thanks{Copyright © 2020 IEEE. Personal use of this material is permitted. However, permission to use this material for any other purposes must be obtained from the IEEE by sending an email to pubs-permissions@ieee.org.}
\thanks{This work was supported in part by the Major Research Plan of the National Natural Science Foundation of China under Grant 61991451 and in part by the Shenzhen Special Fund for the Strategic Development of Emerging Industries under Grant JCYJ20170412170118573.}
\thanks{W.~Xia, Y.~Yang are with Tsinghua Shenzhen International Graduate School, Tsinghua University, China. (Corresponding author: Yujiu Yang)}
\thanks{J.-H.~Xue is with the Department of Statistical Science, University College London, UK.}%
\thanks{J.~Xiao is with Ping An Technology Co., Ltd., Shenzhen, China.}
\thanks{Color versions of one or more of the figures in this article are available online at http://ieeexplore.ieee.org.}
}

\markboth{}{Xia \MakeLowercase{\textit{et al.}}: Domain Fingerprints for NR-IQA}

\maketitle

\IEEEdisplaynontitleabstractindextext

%
\IEEEpeerreviewmaketitle


\begin{abstract}
Human fingerprints are detailed and nearly unique markers of human identity. Such a unique and stable fingerprint is also left on each acquired image. It can reveal how an image was degraded during the image acquisition procedure and thus is closely related to the quality of an image. In this work, we propose a new no-reference image quality assessment (NR-IQA) approach called domain-aware IQA (DA-IQA), which for the first time introduces the concept of domain fingerprint to the NR-IQA field. The domain fingerprint of an image is learned from image collections of different degradations and then used as the unique characteristics to identify the degradation sources and assess the quality of the image. To this end, we design a new domain-aware architecture, which enables simultaneous determination of both the distortion sources and the quality of an image. With the distortion in an image better characterized, the image quality can be more accurately assessed, as verified by extensive experiments, which show that the proposed DA-IQA performs better than almost all the compared state-of-the-art NR-IQA methods.
\end{abstract}

\IEEEpeerreviewmaketitle

\section{Introduction}
\IEEEPARstart{N}{o}-reference image quality assessment (NR-IQA)~\cite{MoorthyB11, MittalMB12, GuZY015, Ma2018Blind} is a fundamental yet challenging task that automatically assesses the perceptual quality of a degraded image without the corresponding reference for comparison. 
It serves as a key component in low-level computer vision, since in many applications it is difficult or even impossible to acquire the non-distorted image as reference to evaluate the quality of a distorted image.
The ill-posed nature of NR-IQA is particularly pronounced for the absence of the prior knowledge about distortion form.

Numerous efforts~\cite{MittalMB12, KangYLD14,BosseMMWS18,YeKKD12, Ye2012Unsupervised, Zhang2015SOM,Moorthy2010A, Saad2012Blind} have been made to extract powerful features to represent images and image degradations.
Traditional hand-crafted feature based methods usually leverage natural scene statistics (NSS)~\cite{Moorthy2010A, Saad2012Blind, MittalMB12} and learning-based metrics~\cite{YeKKD12, Ye2012Unsupervised, Zhang2015SOM}.
For example, Saad~\emph{et al.}~\cite{Saad2012Blind} leverage the statistical features of discrete cosine transform (DCT) for blind image quality assessment. Mittal~\emph{et al.}~\cite{MittalMB12} propose to extract NSS features in the spatial domain to estimate the image quality.
Ye~\emph{et al.}~\cite{YeKKD12, Ye2012Unsupervised} propose codebook representation approaches to predict subjective image quality scores by support machine regression (SVR).
These hand-crafted features, however, lack flexibility and diversity for representing complex diverse degradations.

In recent years, deep learning methods achieve promising results in NR-IQA. 
Due to the extremely limited training datasets, many methods use various data augmentation or multi-task strategies to better exploit the power of Deep Neural Networks (DNNs).
Some methods~\cite{LinW18,Ren2018RAN4IQA,LimKR18} focus on simulating the behavior of Human Visual System (HVS) by Generative Adversarial Networks (GANs). Specifically, Lin \emph{et al.}~\cite{LinW18} propose a discrepancy-guided quality regression network to encode the difference between distorted image and hallucinated reference to make precise prediction. Ren \emph{et al.}~\cite{Ren2018RAN4IQA} propose the restorative adversarial net, \emph{i.e.}, the restorator reconstructs the non-distorted patches while the discriminator tries to distinguish the restored patches from the pristine distortion-free ones, and the evaluator takes the distorted patch and the restored patch as inputs and predicts a perceptually quantified score.
However, these methods treat images of different distortions the same and the discriminative representations of distortions are under-explored. 

\begin{figure}[t]
  \includegraphics[width=\linewidth]{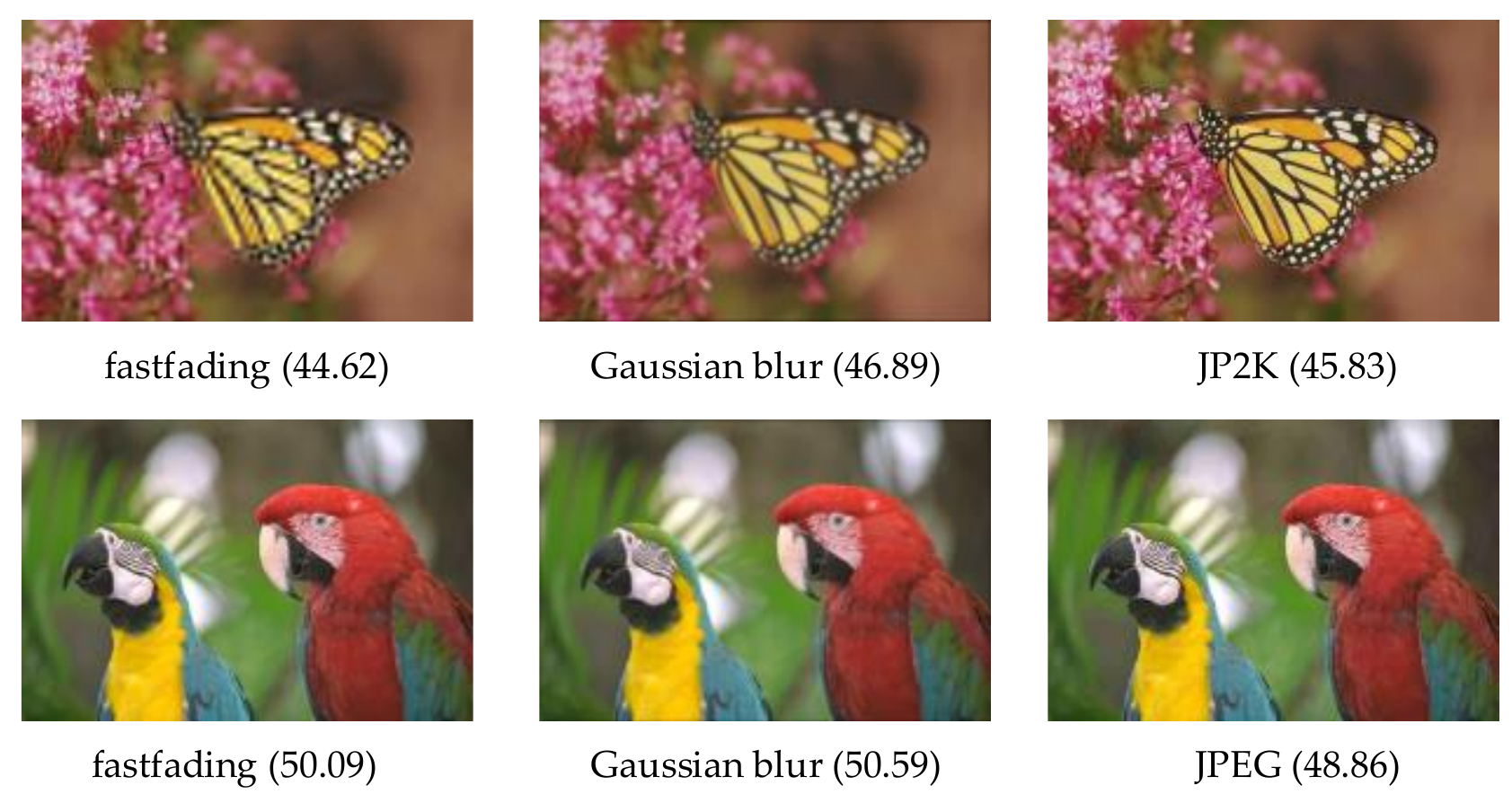}
\caption{\textbf{The two aspects of image quality.} The quality of an image is strongly linked to its distortion type. The degradation arises from the specific distortion, and the image quality depends on the distortion type and level. Higher score denotes worse quality (DMOS, range [0,100]).}
\label{fig:similar_score}
\end{figure}

Different types of distortion change distortion-free images into different distorted versions, leading to significantly different visual perception.
As shown in Figure~\ref{fig:similar_score}, the image quality depends on the distortion type and level. 
Some previous NR-IQA methods~\cite{KangYLD15, Huang19Distortions} try to evaluate the image quality by considering the information of the distortion, but most of them simply use the distortion information by adding a classification to identify the distortion. 

To address the above drawbacks, we propose the domain-aware no-reference image quality assessment (DA-IQA), which exploits the domain fingerprints for image quality assessment. A domain here is defined as the images with the same type and similar levels of distortion. 
Similar to human fingerprints, a unique and stable fingerprint is also left on each acquired image. It can reveal how an image was degraded during the image acquisition procedure and thus is closely related to the quality of an image.
To get domain fingerprints, we design a novel domain-aware architecture to get disentangled representations for different domains. These representations are used as detailed and unique markers to better express particular degradation information, as shown in Figure~\ref{fig:fingerprint}. More specifically, a degraded image can be decomposed as the image content together with the degradation. We disentangle the image content and degradation features from degraded images to more accurately encode degradation information into the image quality assessment framework. The disentangled representations for different domains would be discriminated significantly well in high dimensional feature space.
Furthermore, instead of adding a classification layer as in~\cite{KangYLD15,Huang19Distortions}, benefiting from the design of domain-aware network architecture, DA-IQA can also identify the distortion type of an image simultaneously.

Our contributions are summarized as follows:
\begin{itemize}
\item We propose the domain-aware no-reference image quality assessment (DA-IQA), which for the first time introduces the discriminative disentangled representations for different types of distortions to image quality assessment.
\item Our method can also identify the distortion type of an image, and use the distortion type and quality score to characterize the image quality.
\item Our method achieves superior performance on popular IQA datasets to state-of-the-art methods.
\end{itemize}

\section{Related Work}
\label{sec:Related Work}

\subsection{No-Reference Image Quality Assessment}
\label{sec:biqa}

The existing studies on NR-IQA can be broadly classified into two categories: designing hand-crafted features~\cite{Moorthy2010A, Saad2012Blind, MittalMB12, MoorthyB11} and learning discriminant visual features automatically~\cite{YeKKD12, Ye2012Unsupervised, Zhang2015SOM}. 
The first category of methods typically use a two-stage framework, which performs the distortion identification and the distortion-specific quality estimation accordingly. However, Mittal~\emph{et al.}~\cite{MittalMB12} have shown that such two-stage methods are not superior to the distortion-blind approaches.
The second category of work attempts to learn discriminant visual features automatically without using hand-crafted features. Ye~\emph{et al.}~\cite{YeKKD12, Ye2012Unsupervised} construct a small yet accurate codebook to look up the proper features.
Kang~\emph{et al.}~\cite{KangYLD14, KangYLD15} and Bosse~\emph{et al.}~\cite{BosseMMWS18} adopt deep neural network to extract features from the raw input and perform regression to estimate perceptual scores.
The above NR-IQA methods can be summarized as feature extraction and regression based only on distorted images. 
However, according to the free-energy theory~\cite{Friston06brain}, HVS tends to restore the distorted image before quality assessment. Despite building NR-IQA models based on the free-energy theory, \cite{ZhaiWYLZ12,GuZY015} restore the distorted image with a linear autoregressive model, which is not capable of producing a satisfactory result when the input suffers from high-level distortion and therefore may not be consistent with HVS.
Lin~\emph{et al.}~\cite{LinW18} and Ren~\emph{et al.}~\cite{Ren2018RAN4IQA} simulate the behavior of HVS by using generative adversarial networks (GANs) to generate the corresponding restored counterparts as reference. Ren~\emph{et al.}~\cite{Ren2018RAN4IQA} propose the restorative adversarial net and Lin~\emph{et al.}~\cite{LinW18} propose a discrepancy-guided quality regression network to encode the difference between distorted image and hallucinated reference to make precise prediction.  
However, their methods do not exhibit the capability of disentangling and characterising discriminative latent representations for different degradations, which is one of our key contributions.

\subsection{Representation Disentanglement}
\label{sec:disentangled}
Many recent works on disentangled representation aim to learn an interpretable and transferable representation. For example, Denton \emph{et al.}~\cite{denton2017unsupervised} separate time-independent and time-varying components for long-term video prediction. Some studies~\cite{Xiao2018elegant,Lee2018DRIT,huang2018munit} focus on disentanglement of content and style to achieve multi-domain image translation. It is difficult to define content and style explicitly, and different studies adopt different definitions for their specific tasks. Liu \emph{et al.}~\cite{liu2018unified} propose a unified model that learns disentangled representation for describing and manipulating data across multiple domains. For image restoration, Lu \emph{et al.}~\cite{lu2019unsupervised} disentangle the content and blur features from blurred images. 
Different from~\cite{lu2019unsupervised}, we disentangle the content and discriminative representations of multiple degradations. We use the content features to restore images and use the discriminative representations for NR-IQA. 

\subsection{Image Fingerprint}
\label{sec:fingerprint}
Digital fingerprint is a signature that could be used to identify, track, monitor and monetize images by converting their content into compact digital asset or impression.
Prior digital fingerprint techniques focus on detecting hand-crafted features for device fingerprints~\cite{ChenFGL08}. 
Recently, \cite{MarraGVP19} introduces this concept to the image forensic field and show their application to the GAN source identification.
Based on that, \cite{Yu2019Fingerprints} replaces the hand-crafted fingerprint formulation with a learning-based one, and classify an image as real or GAN-generated by learning GAN fingerprints of different GAN models.

\subsection{Domain-Aware Applications}
The word \emph{Domain-aware} mostly occurs in the NLP applications and its meaning varies from case to case. For example, \cite{Sajal17domain} propose a domain-aware dialog system, which aims to maintain a fluent and natural conversation within the domain as well as during switching of domains. The domain in this case refers to the topic or theme of the conversation. Slightly different, \cite{SahaKS18Conversation} refer domain to a specific field such as retail, travel and entertainment, and introduce the task of multimodal domain-aware conversations.

A \emph{domain} in our work refers to a collection of images with certain degradation. The process of image restoration can be formulated as the translation from the degraded domain to the pristine domain. Thus our work is mostly related to multi-domain image translation. Specifically, \cite{Choi2018stargan} recently proposes a unified model to achieve multi-domain image-to-image translation. \cite{liu2018unified} proposes a model that is able to perform continuous cross-domain image translation and exhibits ability to learn and disentangle desirable latent representations.

In this work, we demonstrate the existence and uniqueness of domain fingerprints, that signify disentangled discriminative representations of different degradations, for image quality assessment tasks.

\begin{figure}[t]
\includegraphics[width=\linewidth]{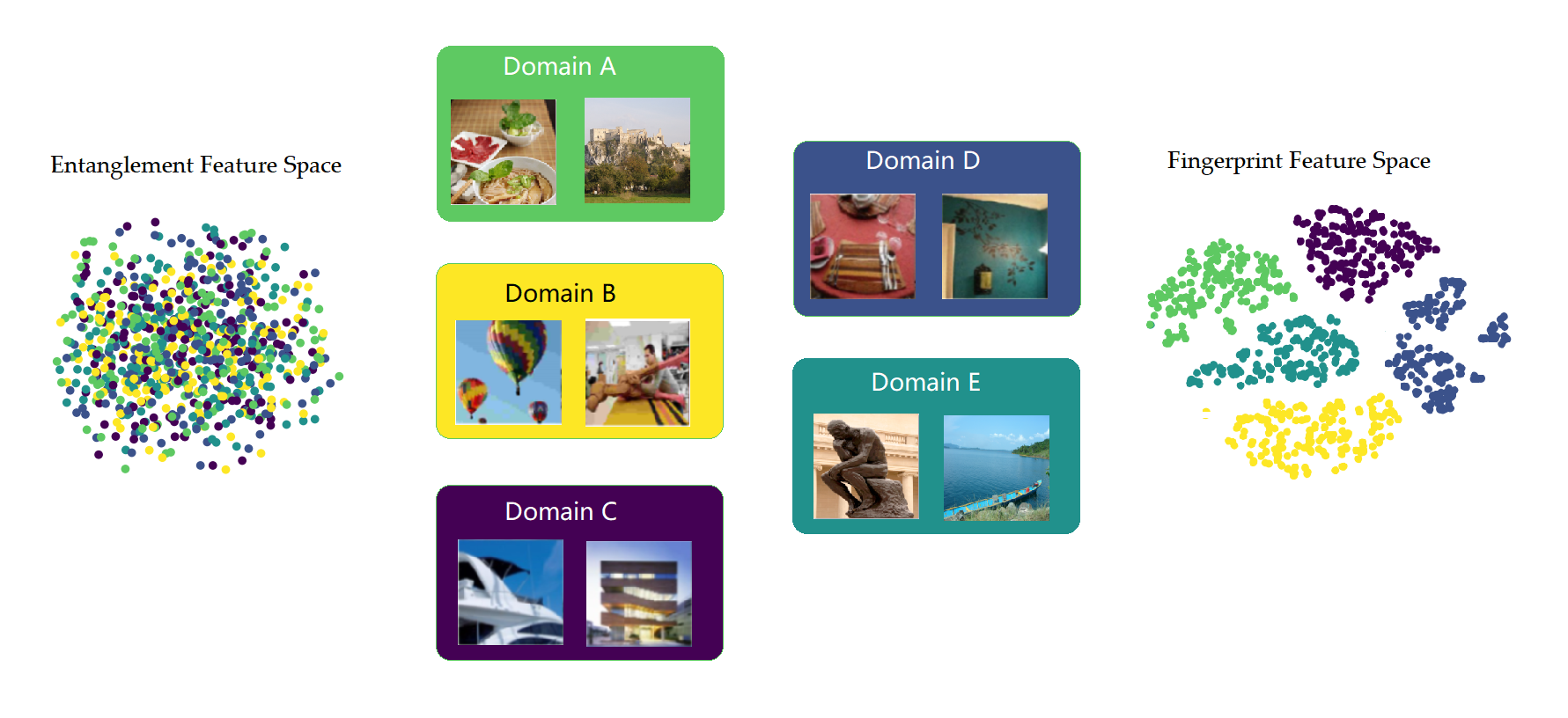}
\caption{\textbf{Illustration of domain fingerprints.} We use t-SNE~\cite{Rauber2016tSNE} for visual comparison between the entangled features (left) and our fingerprint features (right), for degradation representations.}
\label{fig:fingerprint}
\end{figure}

\section{Our Approach}
\label{sec:approach}
\begin{figure*}[t]
  \includegraphics[width=\textwidth]{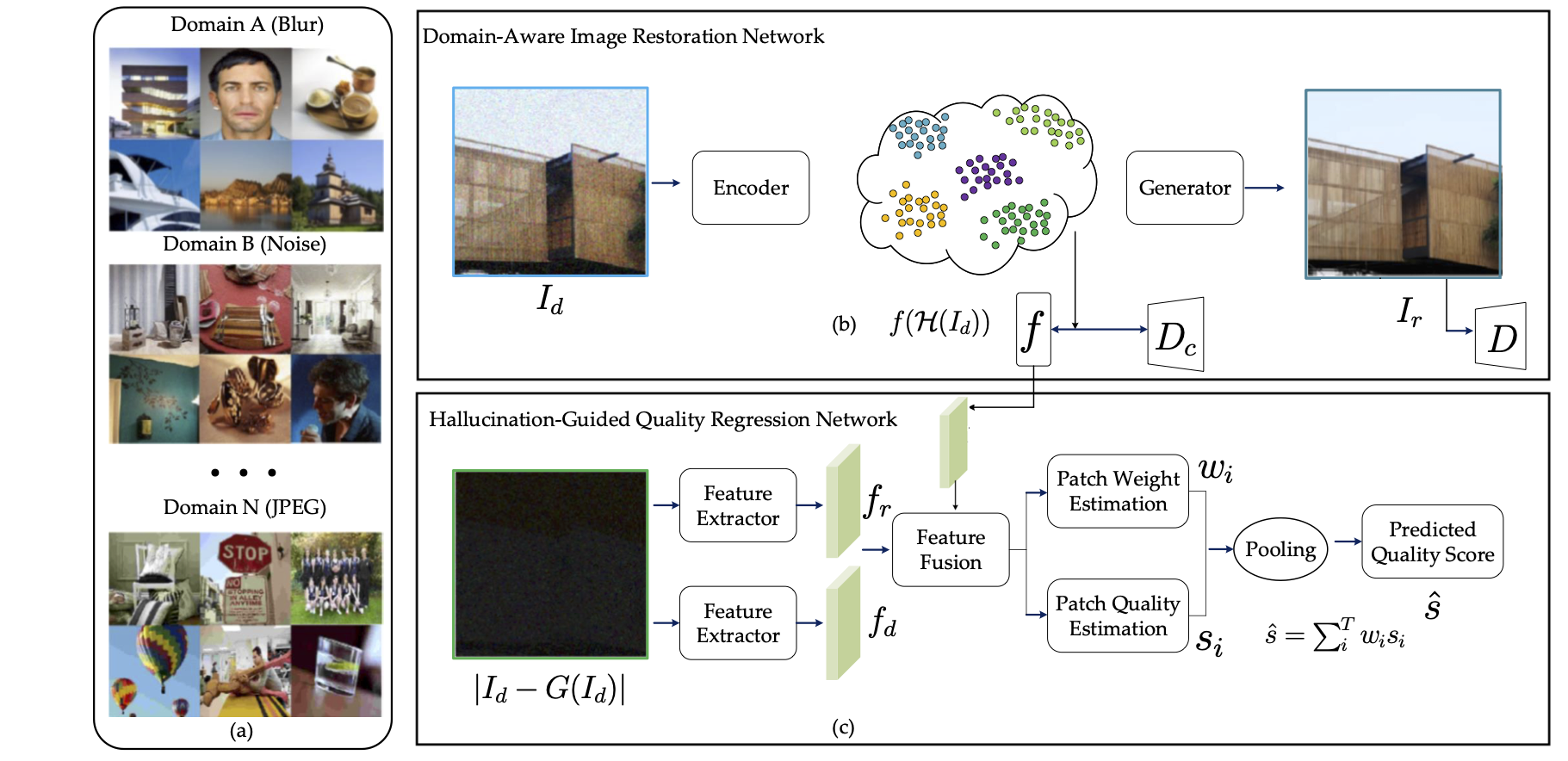}
\caption{\textbf{The framework of the proposed DA-IQA.} Given a dataset that contains several collections of images with different degradations (a), referred to as $n$ domains, we first randomly select a distorted image $I_d$ together with its specific domain label $c$ from a certain domain $\mathcal{D}$, the generator $G$ of Domain-Aware Image Restoration Network (b) aims to separate the distortion $d$ from $I_d$ to get the restoration $I_r$, while the discriminator $D$ tries to distinguish if the input image is real or fake and the domain discriminator $D_c$ recognizes the domain category ($D$ and $D_c$ are omitted from the framework for simplicity).
The original distorted image $I_d$ and its discrepancy map $\left|I_d-G(I_d)\right|$ are fed into the Hallucination-Guided Quality Regression Network (c). The two images are first through feature extractors to get their corresponding features $f_r$ and $f_d$. The two feature extractors share the same weights. Then  $f_r$ and $f_d$ are concatenated with high-level domain-aware distortion representation $f(\mathcal{H}(I_d))$, which is extracted from the restoration network.
The fused feature is regressed to a patch-wise quality and weight estimation. The score of an image is the weighted average of all scores of its patches.}
\label{fig:architectures}
\end{figure*}

\subsection{Problem Formulation}
\label{sec:Problem Formulation}
Given a distorted image $I_d$, our goal is to learn a mapping $f: I_d\rightarrow s$, in which $s \in \mathbb{R}^{+}$ denoted the predicted quality score of $I_d$ and should be consistent with the result of Mean Opinion Score (MOS).
To simulate the Human Visual System (HVS), we first restore degraded contents as reference. Assume in a dataset, there are $n$ domains $\left\{\mathcal{D}_{1}, \mathcal{D}_{2}, \cdots, \mathcal{D}_{n}\right\}$ of images. Each domain represents a collection of images with certain degradation or distortion. For each image $I_d$ in a domain $\mathcal{D}$, it can be represented as the combination of pristine image $I_{gt}$ and a certain distortion $d$, \emph{i.e.}, $I_d=I_{gt} \otimes d$. Our goal is to disentangle the representation $d$ of a distorted image $I_d$ to get the restored image $I_r$ consistent with $I_{gt}$, and use this domain-aware distortion pattern to obtain image quality score $s$ under the supervision of ground truth human visual quality score $s_{gt}$.

\subsection{Overview of the Proposed Approach}
The framework of our proposed DA-IQA is demonstrated in Figure \ref{fig:architectures}. As shown, given several collections of images with different types and levels of degradations, referred to as $n$ domains, an image  $I_d$ is randomly selected from one domain, the generator $G$ tries to produce an image indistinguishable with the real pristine image to the image discriminator $D$. At the same time, the domain discriminator $D_c$ tries to recognize the domain label $c_i$ of representation and also reacts on the generator to further disentangle discriminative features for different degradations. After the convergence of restoration, we further train a regression network by using the high-level domain-aware distorted representation (domain fingerprint) and the obtained restored images $I_r$ to get the desired quality score $s$. The aforementioned processes are implemented by two corresponding modules, the domain-aware image restoration network and the hallucination-guided quality regression network, respectively, as shown in Figure \ref{fig:architectures}.

\subsection{Domain-Aware Image Restoration Network}
\label{sec:DA-Restore}
\begin{figure}[t]
  \includegraphics[width=\linewidth]{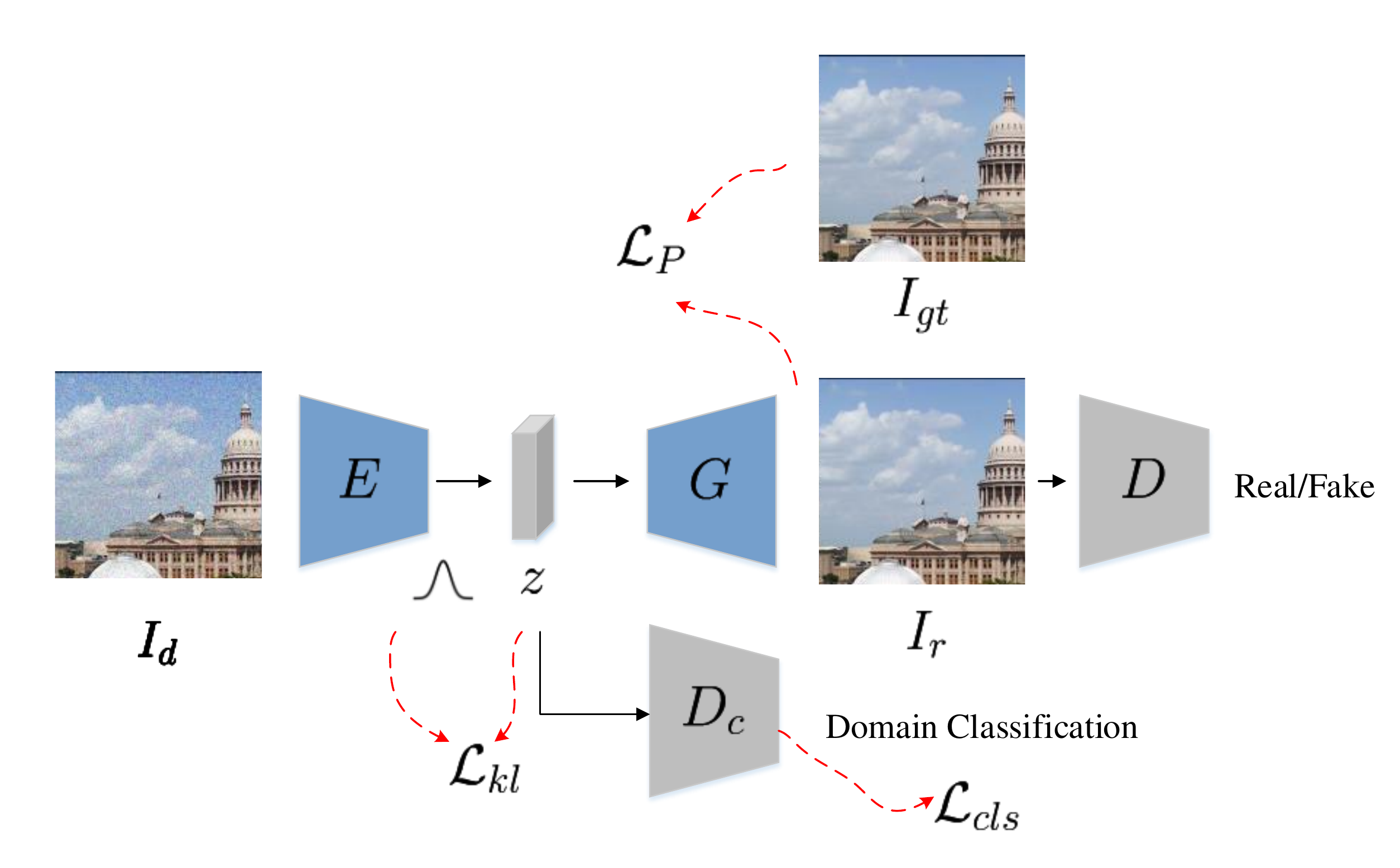}
\caption{\textbf{Overview of our proposed DA-Restore network.} This module takes the distorted image as input and aims to produce its corresponding restoration. Meanwhile, the model tries to disentangle the discriminative representations for different degradations.}
\label{fig:overview}
\end{figure}

The overview of proposed domain-aware image restoration network (DA-Restore) is shown in Figure \ref{fig:overview}. This DA-Restore module takes the distorted image as input and aims to produce the corresponding restoration. Meanwhile, the model tries to disentangle the distorted representation of specific degradation from the content. The restored image could act as a hallucinated reference for the distorted image, which compensates the absence of true reference information and simulates the behavior of the human visual system. Furthermore, due to the design of domain awareness, it determines the distortion type of the input distorted image.

\subsubsection{Latent Feature Loss}
The latent feature loss aims to penalize the latent representations from two aspects. To learn disentangled representation across domains, we use the first term $\mathcal{L}_{kl}=KL(q(z|x) \| p(z))$ to calculate the Kullback-Leibler divergence, which makes the latent code $z$ close to a prior Gaussian distribution $p\left(z\right)$. However, this term alone cannot guarantee the disentanglement of domain-specific information from the latent space, since the generator recovers the distorted-free images simply from the representation $z$ without using any domain information.

To address the above problem and achieve simultaneous training of multiple domains with a single model, we extend the loss by adding a domain classification loss term, to eliminate the domain-specific information from the representation $z$. 
We assign a unique label for each domain, and introduce an auxiliary domain classifier which tries to distinguish the latent representations $z$ from different domains.

Typically, latent representation $z$ can be learned directly from the images through an encoder-decoder architecture. The encoder extracts features from images and the decoder uses these features to reconstruct the images.
The latent feature loss is used to force the encoder to disentangle the content and the degradation information. Without the latent feature loss, the model can still learn a latent representation from the images, but the obtained representation would be entangled as shown in the left-hand panels of Figure \ref{fig:fingerprint} and Figure \ref{fig:visualization}, since no constraints are added on the disentanglement process.

Different from \cite{Choi2018stargan}, which aims to translate facial attributes among domains, image restoration transfers \emph{several} domains of different distortions into one \emph{single} distortion-free domain. Thus, we assign labels to representations instead of images. 
More precisely, we introduce a domain discriminator $D_c$, which takes the latent representation $z$ in domain $c$ as input and aims to distinguish the predicted domain code $v_c$ from its real domain code. In contrast, the encoder $E$ tries to confuse $D_c$ from predicting the correct domain. The second term of latent feature loss, named domain classification loss, can be defined as
\begin{equation}
\mathcal{L}_{E}^{cls}=-\mathcal{L}_{D_c}^{cls}=-\mathbb{E}[-\log P(v_c|E(x_c)],
\end{equation}
where $z=E(x_c)$ with $x_c$ denoting the input image, and $P$ is the probability distribution over domains $c$, which is produced by the domain discriminator $D_c$. Ideally, the latent representation $z$ should contain distortion-free image content information together with disentangled domain-specific degradation information.

\subsubsection{Perceptual Loss}
From our preliminary experiments, we observe some unpleasing artifacts in the restored images. Motivated by~\cite{LinW18}, we add a perceptual loss~\cite{Johnson2016Perceptual} $\mathcal{L}_{{P}}$ between the ground-truth images $I_{gt}$ and the corresponding restored ones $I_r$:
\begin{equation}
\mathcal{L}_{{P}}=\mathbb{E}[\sum_{i} \frac{1}{N_{i}}\left\|\phi_{i}(I_r)-\phi_{i}(I_{gt})\right\|_{1}],
\end{equation}
where $\phi_{i}$ denotes the $i$th layer of the feature maps extracted from the pre-trained VGG-19 network, $N_{i}$ is the number of the selected layers. The features extracted from pre-trained deep networks contain rich semantic information and are widely used in learning-based image generation tasks. The perceptual loss encourages the decoder $\boldsymbol{G}$ to generate images that match ground-truth images perceptually.

Different from the per-pixel loss functions, the perceptual loss compares high level differences such as content and style discrepancies between images, instead of understanding differences at a pixel level. The perceptual loss is a commonly used loss function to provide accurate and photo-realistic results.

\subsubsection{Adversarial Loss}
The VAE architecture tends to generate blurry samples~\cite{Zhao17VAE}, which would not be desirable for practical use. To get satisfactory image restoration from latent representation, we additionally introduce an image discriminator $D$, which also enhances the ability of representation disentanglement from the latent space. We define the objective functions $\mathcal{L}_{D}^{adv}$ and $\mathcal{L}_{G}^{adv}$ for adversarial learning between image discriminator $D$ and image generator $G$ as
\begin{equation}
\begin{aligned}
\mathcal{L}_{D}^{adv} &=\mathbb{E}[\log (D(\hat{x}))]+\mathbb{E}[\log (1-D(x_c))],\\ 
\mathcal{L}_{G}^{adv} &=-\mathbb{E}[\log (D(\hat{x}))],
\end{aligned}
\end{equation}
where $x$ and $\hat{x}$ denote the input image and its restoration, respectively.

\subsubsection{Overall Training Loss}
For stable training, high image quality and considerable diversity, we use the least-squares GAN~\cite{mao2017least} in our experiment. The total training loss functions of the encoder $E$, decoder $G$, image discriminator $D$ and domain discriminator $D_c$ are defined as
\begin{equation}
\begin{aligned}
&\mathcal{L}_{{E}}&=&\lambda_1\mathcal{L}_{kl}+\lambda_2\mathcal{L}_{P}+\mathcal{L}_{E}^{cls},
&\mathcal{L}_{{D_c}}&=&\mathcal{L}_{D_c}^{cls},\\
&\mathcal{L}_{{G}}&=&\lambda_1\mathcal{L}_{kl}+\lambda_2\mathcal{L}_{P}+\mathcal{L}_{G}^{adv},
&\mathcal{L}_{{D}}&=&\mathcal{L}_{D}^{adv},
\end{aligned}
\end{equation}
where $\lambda_1$ and $\lambda_2$ are regularization parameters controlling the importance of losses.

\begin{table*}[t]
\caption{Cross-validation on TID2013 and LIVE. Our proposed method performs better than almost all the other state-of-the-art FR-IQA and NR-IQA methods in terms of both PLCC and SROCC.}
\label{tab:iqa}
\begin{center}
\begin{tabular}{cccccc}
\toprule
\multirow{2}{*}{Category}&\multirow{2}{*}{Method} &\multicolumn{2}{c}{TID2013}&\multicolumn{2}{c}{LIVE}\\
&   & SROCC & PLCC & SROCC & PLCC \\
\hline
\multirow{4}{*}{FR-IQA}
&PSNR &0.889 &0.847 &0.880 &0.805 \\
&SSIM \cite{WangBSS04SSIM} &0.856 &0.867 &0.918 &0.780 \\
&FSIM \cite{ZhangZMZ11FSIM} &\textbf{0.963} &0.932 &\textbf{0.952}& 0.822 \\
&VSI \cite{ZhangSL14} &0.947 &\textbf{0.939} &0.936 &\textbf{0.853} \\
\hline
\multirow{8}{*}{NR-IQA}
&DIVINE \cite{MoorthyB11} &0.855 &0.851& 0.885 &0.853\\
&BLIINDS-II \cite{Saad2012Blind} &0.877 &0.841 &0.931 &0.930 \\
&BRISQUE \cite{MittalMB12}  &0.922 &0.917 &0.940 &0.911 \\
&CORNIA \cite{YeKKD12} &0.903 &0.917 &0.913 &0.888 \\
&CNN  \cite{KangYLD14} &0.903 &0.917 &0.913 &0.888 \\
&CNN++ \cite{KangYLD15} &0.843 &0.804 &0.928 &0.897 \\
&DIQaM-NR \cite{BosseMMWS18} &0.933& 0.909& 0.960& 0.972 \\
&RAN \cite{Ren2018RAN4IQA} &0.948 &\textbf{0.937} &0.972 &0.968 \\
&DA-IQA &\textbf{0.952} & 0.929&\textbf{0.977} &\textbf{0.975}  \\
\bottomrule
\end{tabular}
\end{center}
\end{table*}

\subsection{Hallucination-Guided Quality Regression Network}
\label{sec:DA-IQA}
As shown in Figure \ref{fig:architectures}, the DA-Restore module generates the residual image for restoration. This residual image is different from the concept of \emph{error map} in FR-IQA, which represents pixel-wise error between the distorted image and the reference. To emphasize the difference, we refer to the residual image as \emph{discrepancy map} and use the original distorted image together with the discrepancy map as input to acquire the desired quality score. The NR-IQA problem is now formulated as a regression by solving
\begin{equation}
\hat{\theta}=\arg \min _{\theta} \frac{1}{N} \sum_{i=1}^N \mathcal{L}(\mathcal{R} ({I_d^i, \left|I_d^i-G(I_d^i)\right|}), s^i_{gt}),
\end{equation}
where $\mathcal{R}(\cdot)$ denotes the regression network for predicting the quality score, $\left|I_d^i-G(I_d^i)\right|$ is the discrepancy map generated by the DA-Restore module $G$, and $s_{gt}$ represents the ground truth score.

Previous studies have shown that the quality results obtained by methods based on HVS are greatly affected by the eligibility of the restoration~\cite{LinW18,Ren2018RAN4IQA}. An unqualified hallucination as the reference would introduce a large bias by deteriorating the gap between the distorted image and the restored one. To alleviate this drawback and stabilize the quality regression, we fuse the discrepancy information together with the high-level information from the generative network as with~\cite{LinW18}. Different from their work, ours uses the domain-specific distortion information.

Assume $G$ has been trained. As shown in Figure \ref{fig:architectures}, features of the original distorted image and its \emph{discrepancy map}, $f_r$ and $f_d$, are extracted by a CNN and are concatenated with the high-level domain-aware distortion representation $f(\mathcal{H}(I_d))$, which is extracted from the restoration network. $\mathcal{H}(I_d)$ denotes the feature maps of the distorted image, $f(\cdot)$ is a linear projection to ensure that the dimensions of $\mathcal{H}$ and $\mathcal{K}$ are equal, and $\mathcal{K}$ represents the feature maps that concatenate with $\mathcal{H}$.
Then the fused feature is regressed to a patch-wise estimation of quality and weight estimation, denoted by $s_{i}$ and $w_{i}$, respectively, and $w_{i}$ is treated as the relative importance for each patch $i$.

The loss function of the patch quality estimation would be formulated as
\begin{equation}
    \mathcal{L}_R=\frac{1}{T}\sum_{i=1}^{T}\left\|s_i-s_{gt}\right\|_1,
\end{equation}
where $s_i=\mathcal{R}_2(f(\mathcal{H}(I_d))\oplus\mathcal{K}(I_d, \left|I_d^i-G(I_d^i)\right|))$, $\mathcal{R}_2$ is the fully connected layers of $\mathcal{R}(\cdot)$, $\oplus$ denotes the concatenation operation, and $T$ is the number of patches.

We assign different weights for the respective patches and use the normalized weights to estimate the quality score of the whole image. The reason of integrating a branch of weight estimation is that simply averaging all local quality scores does not consider the effect of spatial variance of relative image quality and perceptual relevance of local quality. The two branches are parallel and share the same dimension. The weight $w_i$ of each input patch $i$ is calculated by activating the output of weight estimation branch $w^{\prime}$ through a ReLU and adding a small stability term:
\begin{equation}
    w_i=\max(0,w^{\prime})+\epsilon.
\end{equation}

The final image quality score $\hat{s}$ is thus
\begin{equation}
    \hat{s}=\sum_i^{T} w_i s_i.
\end{equation}

\subsection{Network Architecture}
For the domain-aware image restoration network, inspired by recent image translation studies, our generators follow an encoder-decoder architecture similar to that in~\cite{liu2018unified}.
Given depth $d=5$, the $i$th layer of Encoder $E$ operates on $4\times4$ spatial regions with a stride 2 and produces a feature map with size of $\{64\times2^{i-1}\}^d_{i=1}$, \emph{i.e.},16, 128, 256, 512, 1024, respectively. Each convolutional layer is followed by Instance Normalization~\cite{ulyanov2016instance}, and Leaky ReLU are utilized.
Generator $G$ follows a reversed symmetrical architecture of $E$. Residual learning is adopted because it has been shown effective for image processing tasks and helpful on convergence. That is, the generator only learns the difference between the input image and the ground truth image. The domain discriminator $D_c$ (discriminator with an auxiliary domain classifier) is built on top of the discriminator $D$.

For the hallucination-guided quality regression network, the domain-aware distortion representation is extracted from the well-trained restoration network and acts as compensation for IQA. The hallucination-guided quality regression network takes the distorted patch and the corresponding restored residual patch as input. Feature representations $f_r$ and $f_d$ are extracted by the same layers and fused with obtained feature map $z$. The fused feature vector is then fed into two branches for estimating perceptual score $s_i$ and weight $w_i$, respectively.

\section{Experiments}
\label{sec:experiments}
In this section, we conduct several experiments to test the performance of our proposed method on various IQA benchmarks. We pre-train DA-IQA on Waterloo Exploration~\cite{MaDWWYLZ17} and perform cross validation on TID2013~\cite{PonomarenkoJILEAVCCBK15} and LIVE~\cite{SheikhSB06}.

\subsection{Datasets and Evaluation Protocols}
\label{sec:experiment setting}

\subsubsection{TID2013} TID2013 includes 25 distortion-free reference images and 3000 distorted images. These images are created from references with 24 types and 5 levels of distortions, ranging from additive Gaussian noise to Chromatic aberrations. Every image is annotated with Mean Opinion Scores (MOS), which is produced by several observers in subjective tests.  The obtained MOS has to vary from 0 to 9 and its larger values correspond to better perceptual quality. Its wide range makes it one of the most comprehensive IQA databases.

\subsubsection{LIVE} LIVE consists of 29 reference images and 779 distorted samples with 5 distortion types including Fast Fading, Gaussian Blur, White Noise, JPEG Compression and JP2K Compression. Each image is provided with Differential Mean Opinion Scores (DMOS), ranging from 0 to 100. Lower DMOS means higher perceptual quality. DMOS value of zero indicates the image is distortion-free.

\subsubsection{Waterloo Exploration} Waterloo Exploration contains 4744 pristine natural images and 94880 distorted images. The distorted images are generated by MATLAB with four distortion types and five levels. Compared to TID2013 and LIVE, Waterloo Exploration has much larger amounts of images, thus it also has a great diversity of image content. The four types, \emph{i.e.}, JPEG Compression, JP2K Compression, Gaussian Blur and White Noise, are also considered the most common distortion types and are covered both in TID2013 and LIVE. 

\subsubsection{Evaluation Metrics}
Following most previous works~\cite{KangYLD14,BosseMMWS18, Ren2018RAN4IQA}, the performances on above datasets are evaluated by two common metrics for model evaluation: the Linear Correlation Coefficient (LCC) and the Spearman's Rank Order Correlation Coefficient (SROCC). LCC is a measure of the linear correlation between the ground-truth and model prediction, which is defined as
\begin{equation}
    \text{LCC}=\frac{\sum_{i=1}^{N}\left(y_{i}-\overline{y}_{i}\right)\left(\hat{y}_{i}-\overline{\hat{y}}_{i}\right)}{\sqrt{\sum_{i=1}^{N}\left(y_{i}-\overline{y}_{i}\right)^{2}} \sqrt{\sum_{i=1}^{N}\left(\hat{y}_{i}-\overline{\hat{y}}_{i}\right)^{2}}},
\end{equation}
where $\overline{y}_{i}$ and $\overline{\hat{y}}_{i}$ denote the means of the ground truth and predicted score, respectively.
SROCC measures the prediction monotonicity between the ground-truth and model prediction, which could be formulated as
\begin{equation}
\text{SROCC}=1-\frac{6 \sum_{i=1}^{N}\left(v_{i}-p_{i}\right)^{2}}{N\left(N^{2}-1\right)},
\end{equation}
where $N$ represents the number of distorted images, $v_{i}, p_{i}$ are the positions of $\hat{y}_{i}, \overline{\hat{y}}_{i}$ in the ranking sequences. The SROCC measures the monotonic relationship between the ground-truth and estimated IQAs while the PLCC is a measure of the linear correlation between the ground-truth and predicted quality scores. Higher SROCC score means higher monotonicity and higher PLCC score represents higher linear correlation, between the ground-truth and predicted quality scores.

\begin{figure*}[th]
    \includegraphics[width=\linewidth]{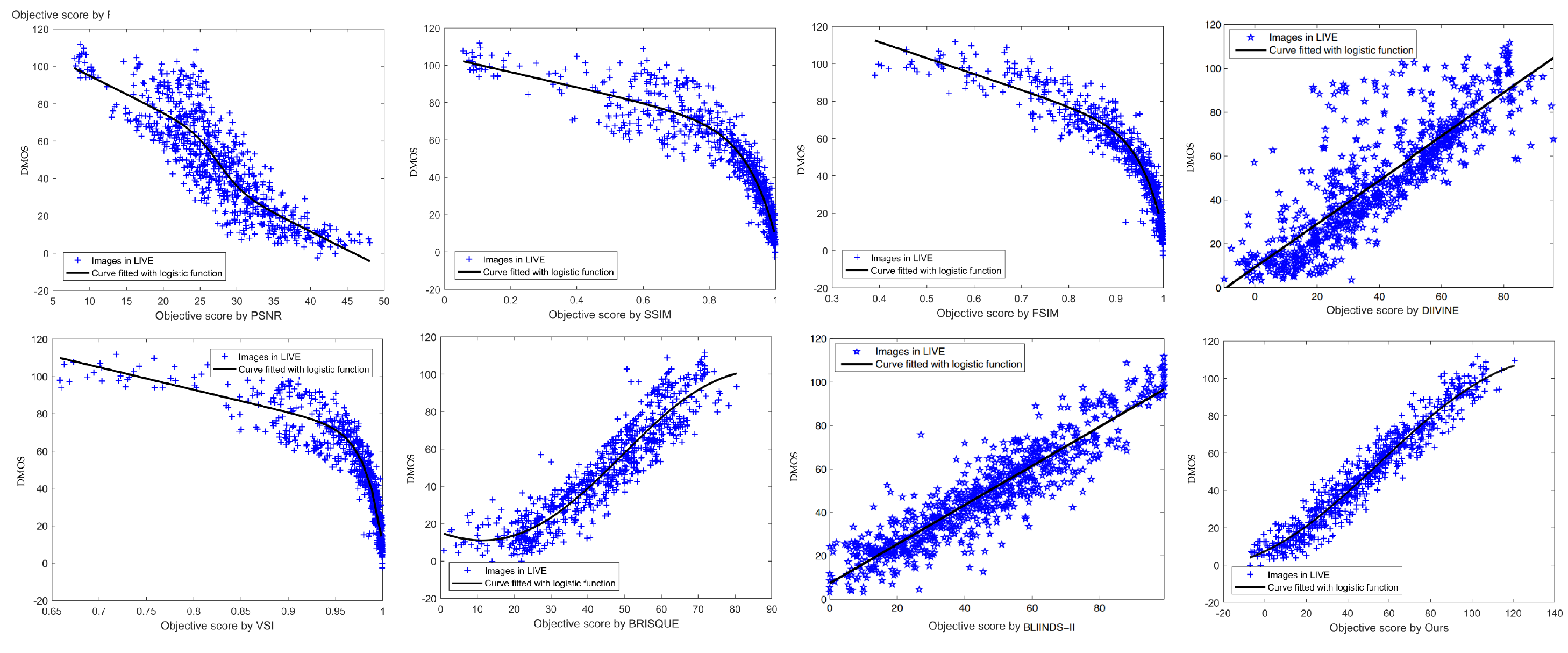}
    \caption{The scatter plot of MOS values with respect to objective values on the LIVE dataset. The blue ``+'' represents one distorted image and the black curves are obtained in the curve fitting process. The blue ``+'' of our method gather evenly and close to the black curve and the curve also exhibits almost a straight line, which manifests the better correlation between the scores given by our method and the subjective judgements for the image quality.}
\label{fig:analysis}
\end{figure*}

\subsection{Implementation Details}
All the training images are randomly sampled from the original images of size $256 \times 256$ with stride $96$. We train our model with Pytorch on the GeForce GTX 1080Ti with a batch size of $32$. We apply Adam~\cite{Kingma2014Adam} solver with parameters of learning rate $0.0002$, $\beta_{1}=0.5, \beta_{2}=0.999$. For domain-aware image restoration, the weights $\lambda_1$, $\lambda_2$ are set as $5$ and $100$, respectively. For hallucination-guided quality regression, we split the TID2013 and LIVE datasets into 6:2:2 for training, validation and test, respectively.
Considering that perceptual scores are not available in Waterloo Exploration, we provide a score and weight for every distorted image patch by performing FSIM \cite{ZhangZMZ11FSIM}, which is one of the state-of-the-art FR-IQA metrics. These scores and weights serve as the ground truth label in the following training process as in \cite{Ren2018RAN4IQA}. During the testing of  image quality assessment, we extract overlapped patches from each test image at a fixed stride and calculate the weighted scores of all patches as the final quality score.

\subsection{Cross Validation on TID2013 and LIVE}
\label{sec:four types}
We firstly train the domain-aware image restoration module on Waterloo Exploration, then train the hallucination-guided quality regression on TID2013 and LIVE to perform cross validation, respectively. Since Waterloo Exploration contains only four distortion types, we train and test DA-IQA on these following types: Gaussian Blur, White Noise, JPEG and JP2K, following the same setting as in~\cite{Ren2018RAN4IQA}. 
As shown in Table~\ref{tab:iqa}, the proposed DA-IQA performs better than almost all the other state-of-the-art FR-IQA and NR-IQA methods in terms of both PLCC and SROCC.
Furthermore, our model outputs simultaneously quality estimation and distortion identification. As shown in Table~\ref{tab:distortion identification}, our method also wins the distortion identification task compared with other distortion-identification IQA methods~\cite{Ma2018Blind,KangYLD15}.
Table~\ref{tab:confusion_matrices_on_LIVE} shows the confusion matrices produced by our method on the LIVE dataset. The column and the raw contain ground truths and predicted distortion types, respectively.

For visualization, we also provide the scatter plot of subjective MOS values with respect to objective values on the LIVE database in Figure \ref{fig:analysis}, in which we denote the distorted images with blue ``+'' and the black curves are obtained in the curve fitting process as in~\cite{Liu2019Camera}. One can see the blue ``+'' of our method gather evenly and close to the black curve and the curve also exhibits almost a straight line, which manifests the better correlation between the scores given by our method and the subjective judgements for the image quality.

\begin{table}[ht]
\caption{Performance of distortion identification on 5 common distortions of LIVE and 24 types of distortion on TID2013.}
\label{tab:distortion identification}
\begin{center}
\begin{tabular}{l|cc}
\toprule
\multirow{2}{*}{Method} &\multicolumn{2}{c}{Accuracy}\\
 &LIVE& TID2013\\
 \hline
CNN++ \cite{KangYLD15}  &0.925 &0.819\\
MEON \cite{Ma2018Blind} &0.912  &0.859\\
DA-IQA               &\textbf{0.942} &\textbf{0.937}\\
\bottomrule
\end{tabular}
\end{center}
\end{table}

\begin{table}[t]
\caption{The confusion matrices produced by our method on the LIVE dataset. The column and the raw contain ground truths and predicted distortion types, respectively.}
\label{tab:confusion_matrices_on_LIVE}
\begin{center}
\begin{tabular}{ccccccc}
\toprule
&Category & JP2K & JPEG & WN & BLUR & Pristine \\
\hline
\multirow{5}{*}{LIVE}
&JP2K       &\textbf{0.915} &0.010&0.000 &0.023&0.032 \\
&JPEG       &0.048 &\textbf{0.919}&0.000 &0.022&0.011\\
&WN         &0.000 &0.000 &\textbf{1.000}& 0.000 &0.000 \\
&BLUR       &0.059 &0.007&0.000 &\textbf{0.926}&0.008 \\
&Pristine   &0.007 &0.013&0.000 &0.034&\textbf{0.950}\\
\bottomrule
\end{tabular}
\end{center}
\end{table}

\subsection{Results on More Distortion Types}
We show results on four distortions in Section~\ref{sec:four types}, to demonstrate the scalability of our domain-aware framework when handling more distortion types, we further train and test on the full TID2013 dataset. For pre-training, as in the previous studies~\cite{liu2017rankiqa,Ma2018Blind}, we reproduce 17 out of a total of 24 distortion types in TID2013 and apply them to the 840 high-quality images. For the distortions we did not reproduce (\emph{i.e.}, \#3, \#4, \#12, \#13, \#20, \#21, \#24), we fine-tune from other distortions. We follow the experimental setting as used in~\cite{liu2017rankiqa} and the average SROCC of all the experiments performed 10 times is reported in Table~\ref{tab:srocc_on_tid2013}, where \textit{ALL} means testing all distortions together. From Table~\ref{tab:srocc_on_tid2013}, several patterns can be observed. First, our proposed DA-IQA outperforms previous models by a clear margin, which strongly demonstrates the effectiveness of domain fingerprints. Second, the DA-Restore has learned domain-aware features, which are fed into the subsequent module to get the desired score. 
To some extent, image restoration and image quality assessment are inherently consistent. The features learned during the restoration can not only be used to reconstruct distort-free images but also contain the degradation type and level information of given images that can be further processed to get the quality score. However, it is hard to obtain satisfactory results by directly training a deep network on IQA data due to the data scarcity. Thus we pre-trained the DA-Restore module on other large-scale high-quality datasets to learn the domain-aware features of different degradations (domain fingerprints) and fine-tuned the quality regression module on accurate IQA scores.

\begin{table*}[t]
\caption{Performance evaluation (SROCC) on the entire TID2013 dataset.}
\label{tab:srocc_on_tid2013}
\begin{center}
\begin{tabular}{c|ccccccccccccc}
\toprule
Method&\#01& \#02& \#03 &\#04& \#05 &\#06& \#07& \#08& \#09 &\#10 &\#11& \#12& \#13 \\
\hline
BLIINDS-II \cite{Saad2012Blind} &0.714 &0.728 &0.825 &0.358 &0.852 &0.664 &0.780 &0.852 &0.754 &0.808 &0.862 &0.251 &0.755\\
BRISQUE \cite{MittalMB12}  &0.630 &0.424& 0.727 &0.321 &0.775 &0.669 &0.592 &0.845 &0.553 &0.742 &0.799 &0.301 &0.672\\
CORNIA \cite{YeKKD12} &0.341 &-0.196 &0.689& 0.184 &0.607 &-0.014 &0.673 &0.896 &0.787 &0.875 &0.911 &0.310 &0.625\\
RankIQA &0.667 &0.620 &0.821 &0.365 &0.760 &0.736 &0.783 &0.809 &0.767 &0.866 &0.878 &0.704 &0.810\\
DA-IQA &0.903 &0.801 &0.905 &0.714 &0.891 &0.879 &0.921 &0.912 &0.897 &0.919 &0.925 &0.609 &0.651\\
\hline
Method &\#14 &\#15 &\#16 &\#17 &\#18 &\#19 &\#20 &\#21 &\#22 &\#23 &\#24 &ALL&\\
\hline
BLIINDS-II \cite{Saad2012Blind} &0.081 &0.371 &0.159 &-0.082 &0.109 &0.699 &0.222 &0.451 &0.815 &0.568 &0.856 &0.550&\\
BRISQUE \cite{MittalMB12} &0.175 &0.184 &0.155 &0.125 &0.032 &0.560 &0.282 &0.680 &0.804 &0.715 &0.800 &0.562&\\
CORNIA \cite{YeKKD12} &0.161 &0.096 &0.008 &0.423 &-0.055 &0.259 &0.606 &0.555 &0.592 &0.759 &0.903 &0.651&\\
RankIQA &0.597 &0.622 &0.268 &0.613 &0.662 &0.619 &0.644 &0.800 &0.779& 0.629 &0.859 &0.780&\\
DA-IQA &0.477 &0.695 &0.438 &0.674 &0.709 &0.852 &0.713 &0.897 &0.808 &0.774 &0.868 &0.828&\\
\bottomrule
\end{tabular}
\end{center}
\end{table*}

\subsection{Feature Disentanglement Visualization}
\label{sec:analysis}
\begin{figure}[ht]
  \includegraphics[width=0.5\textwidth]{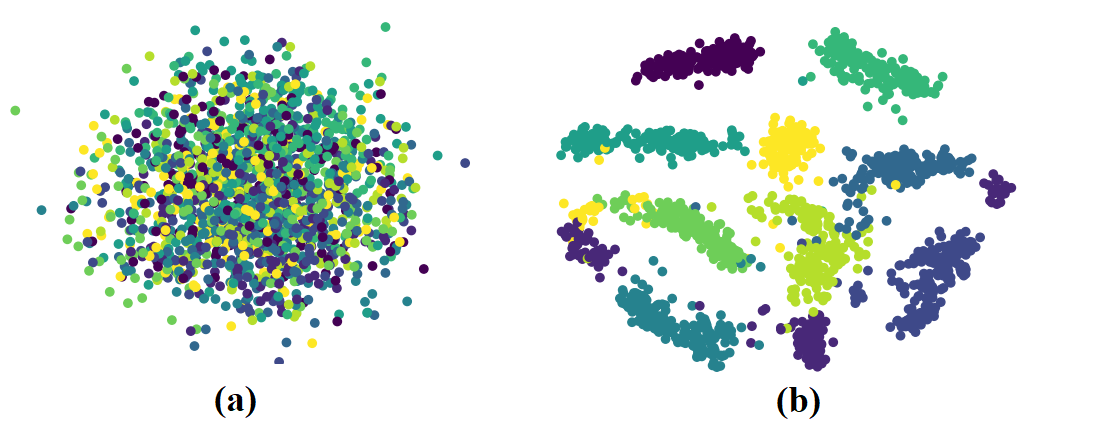}
\caption{Feature visualization (a) without and (b) with domain-aware mechanism. Better view in color.}
\label{fig:visualization}
\end{figure}

To demonstrate the ability for disentanglement and transferability of learned features, many works~\cite{Hoffman2014lsda,liu2018unified} conduct feature visualization with t-SNE~\cite{Rauber2016tSNE}. Similarly, we simultaneously perform feature disentanglement of distortions at different domains and show the results in Figure \ref{fig:visualization}. To be more specific, we trained the DA-Restore module alone with the DIV2K dataset \cite{Agustsson_2017_CVPR_Workshops} aiming to restore images of multiple degradations in one model. The DIV2K \cite{Agustsson_2017_CVPR_Workshops} is often used in the image restoration task as the augmented training dataset.
Each color indicates a different domain, \emph{i.e.}, noisy images with $\sigma$ 15, 25, 30, 50 and 70, JPEG compressed images with quality factor 10 and 20, and low-resolution images with factor 2, 3 and 4. As shown, features of images from different domains are discriminated significantly well.

\subsection{Fingerprint Visualization}
\label{sec:fingerprint_vis}
In Section \ref{sec:disentangled} and Section \ref{sec:analysis}, we describe how to learn and analyze the domain fingerprint which are \textbf{implicitly} represented in the feature domain. 
To help better understand the domain fingerprints, following the similar spirit to~\cite{MarraGVP19,Yu2019Fingerprints}, we \textbf{explicitly} represent them in the image domain by introducing the image fingerprint $F_{im}^I$ and the model fingerprint $F_{mod}^d$. 
The image fingerprints are defined as the reconstruction residual, and the domain fingerprints are learned simultaneously from each domain.
The response is simply defined as pixel-wise multiplication of two normalized images $F_{im}^I \odot F_{mod}^d$.

\begin{figure}[ht]
\begin{center}
\includegraphics[width=0.5\textwidth]{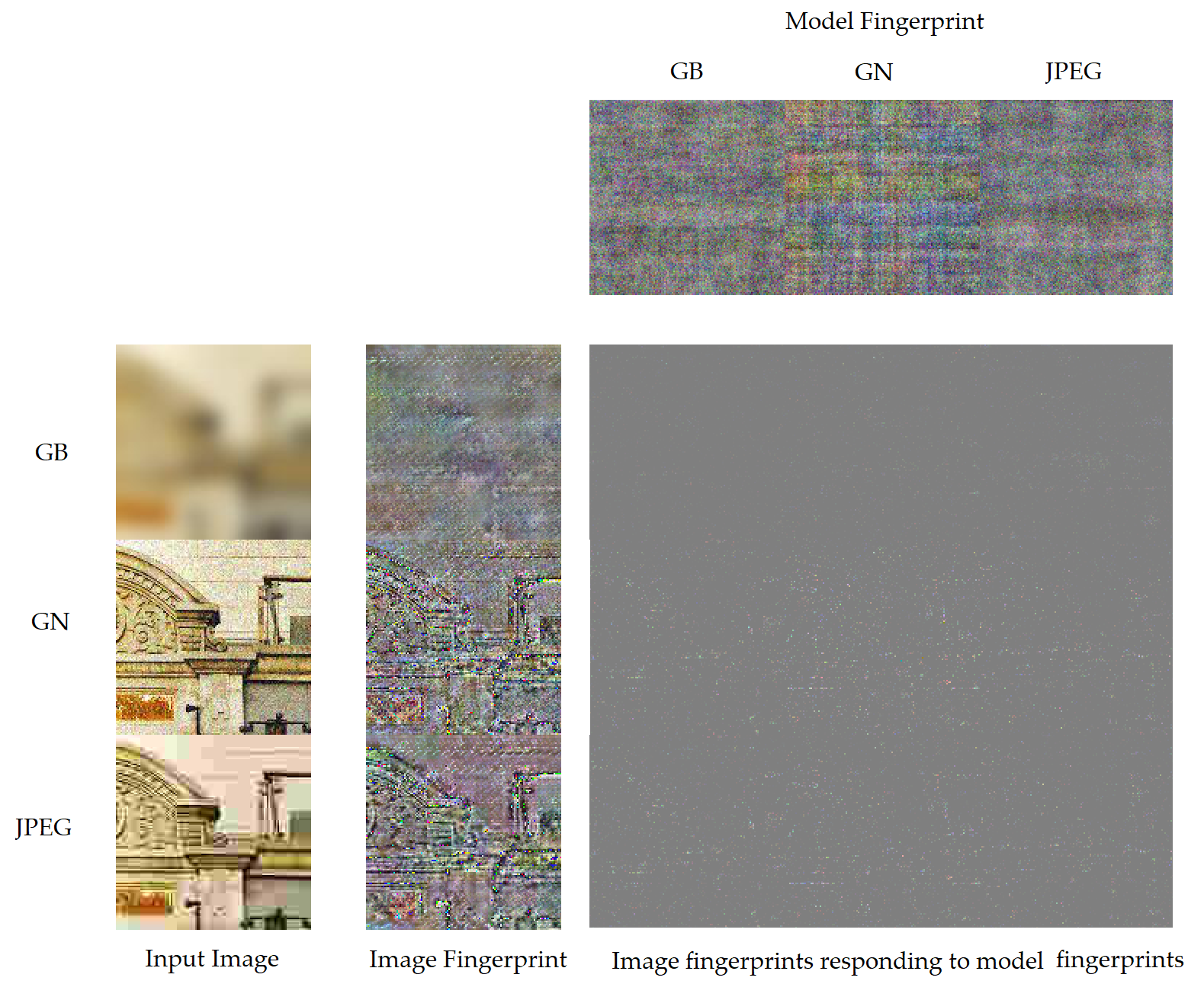}
\end{center}
\caption{Visualization of model and image fingerprint samples. Their pairwise interactions are shown as the confusion matrix. It is simply defined as pixel-wise multiplication of two normalized images.}
\label{fig:fingerprint_vis}
\end{figure}

As shown in Figure~\ref{fig:fingerprint_vis}, we can see that images from different degradations possess different fingerprint patterns. 
Their pairwise interactions are shown as the confusion matrix.
Different from the confusion matrix in Table~\ref{tab:confusion_matrices_on_LIVE}, which compares final predicted distortion types and ground truth labels, this confusion matrix is image fingerprint responding to model fingerprint.
Ideally, the image fingerprints should maximize responses only to their own domain fingerprints. The results in Figure~\ref{fig:fingerprint_vis} are not visually and perceptually obvious.
This can be due to two reasons:
a)~Incomplete feature disentanglement. 
As emphasized in~\cite{xia2019domain}, during feature disentanglement, the degradation information extracted from a single image contains some more information rather than only generalized fingerprint of image collections. 
In the training phase, the model may extracts some content features as degradation fingerprints incorrectly, which means that the fingerprints still contain some general shared image content.
b)~The reconstruction error. 
We define the reconstruction residual as the image fingerprint. It is convenient for visualizing and understanding implicit features. 
However, the reconstruction results of our DA-Restore module could not treat to the same extent as the final restoration. Our goal is to learn and analyze useful feature for image quality assessment instead of restore images. 
Nonetheless, an improved strategy for thorough disentanglement~\cite{xia2019domain} and reconstruction is our future work.

\subsection{Ablation Study}
\label{sec:ablation}
To demonstrate the efficacy of the key components of our method for the performance, we conduct several ablation experiments on TID2013, in which we remove perceptual loss, adversarial learning, high-level semantic fusion, or domain classification and test the performance of the remaining framework by comparing both SROCC and LCC results, as shown in Table \ref{tab:ablation}.

\begin{table}[ht]
\caption{Ablation Experiment on TID2013. ``w/o M'' means our model without component M.}
\begin{center}
\begin{tabular}{l|cc}
\toprule
\multirow{2}{*}{Ablation} &\multicolumn{2}{c}{TID2013}\\
 &PLCC & SROCC\\
 \hline
Proposed (Ours)                 &\textbf{0.929} &\textbf{0.952}\\
w/o Domain Classification       &0.913 &0.934 \\
w/o Semantic Fusion             &0.917  &0.902 \\
w/o Perceptual Loss             &0.883 &0.876 \\
w/o Adversarial Learning        &0.864 &0.859 \\
\bottomrule
\end{tabular}
\end{center}
\label{tab:ablation}
\end{table}

\subsubsection{Domain Classification}
To show how ``domain classification'' contributes to the performance, we remove the domain classification mechanism. Domain classification mechanism is a crucial component to make our approach ``domain-aware''. As illustrated in Figure \ref{fig:visualization}, without domain classification, the representation disentanglement would be only separation of content and degradation. The representations of different degradations are still entangled and no discriminative features of certain degradation are learned. The results in Table \ref{tab:ablation} show that the discrimination of different degradations can actually boost the performance of the IQA task.

\subsubsection{High-Level Semantic Fusion}
The two aforementioned mechanisms contribute to the IQA performance by indirectly improving the quality of restored reference. When the restoration is unqualified, a large bias would be introduced and lead to deterioration of the gap between the distorted image and the restored one. Thus we design high-level semantic fusion to alleviate this drawback and stabilize the quality regression. To show its impact, we remove the fusion module, and the ablation results shown in Table \ref{tab:ablation} demonstrate the validity.

\subsubsection{Perceptual Loss}
We first evaluate the effect of perceptual loss. The ablated model obtains an obvious performance decline by removing the perceptual loss since such a loss helps to restoration at the training process.

\subsubsection{Adversarial Learning}
To explore how adversarial learning contributes to the restoration performance, we further evaluate the model without image discriminator and adversarial loss. Removal of adversarial learning leads to significant performance decline since the discriminator no longer propelled the generator.

\section{Conclusion}
\label{sec:conclusion}
In this paper, we propose the domain-aware no-reference image quality assessment (DA-IQA). 
The proposed DA-IQA leverages domain fingerprints for image quality assessment. These domain fingerprints reveal the disentangled discriminative feature representations of different degradations.
Benefiting from the design of domain-aware network architecture, our method is also able to identify the distortion type of an image, and use the determined distortion type and quality score to characterize the image quality. Experiments on various standard IQA datasets have shown its superiority over state-of-the-art IQA methods.

\bibliographystyle{IEEEtran}
\bibliography{reference}

\end{document}